# Degenerate solutions to the massless Dirac and Weyl equations and a proposed method for controlling the quantum state of Weyl particles

Georgios N. Tsigaridas[1,*], Aristides I. Kechriniotis[2], Christos A. Tsonos[2] and Konstantinos K. Delibasis[3]

[1]Department of Physics, School of Applied Mathematical and Physical Sciences, National Technical University of Athens, GR-15780 Zografou Athens, Greece

[2]Department of Physics, University of Thessaly, GR-35100 Lamia, Greece

[3]Department of Computer Science and Biomedical Informatics, University of Thessaly, GR-35131 Lamia, Greece

[*]Corresponding Author. E-mail: gtsig@mail.ntua.gr

**Abstract**

In a recent work, we have shown that all solutions to the Weyl equation and a special class of solutions to the Dirac equation are degenerate in the sense that they remain unaltered under the influence of a wide variety of different electromagnetic fields. In this study, our previous work is significantly extended, providing a wide class of degenerate solutions to the Dirac equation for massless particles. The electromagnetic fields corresponding to these solutions are calculated and examples regarding both spatially constant electromagnetic fields and electromagnetic waves are also provided. Furthermore, some general solutions to the Weyl equation are presented, and the corresponding electromagnetic fields are calculated. Based on these results, a method for fully controlling the quantum state of Weyl particles using appropriate electromagnetic fields is proposed. Finally, the transition from degenerate to non-degenerate solutions as the particles acquire mass is discussed.

1. Introduction

In recent years it has been shown that in certain materials, such as graphene sheets and Weyl semimetals, ordinary charged particles can collectively behave as massless [1-6], offering a wide range of opportunities for novel technological applications [7-9]. Therefore, an in-depth understanding of the fundamental properties of these particles and especially their interactions with electromagnetic fields, is highly desirable for the efficient utilization of these materials in novel nanoelectronic and nanophotonic components and devices.

In the present work, we utilize a particularly important result published in a recent article by our group [10] to thoroughly study the electromagnetic interactions of these exotic massless particles. In more detail, in [10] we showed that all solutions to the Weyl equation, as well as a special class of solutions to the Dirac equation regarding massless particles, are degenerate, in the sense that they remain unaltered in the presence of a wide variety of electromagnetic fields.

In this study our previous results are extended considerably, providing, in Section 2, a general class of degenerate solutions to the Dirac equation for massless particles. The electromagnetic fields corresponding to these solutions are calculated and examples of both spatially constant electromagnetic fields and electromagnetic waves are discussed. More specifically, we have shown that the state of massless Dirac and Weyl particles is not affected by the presence of a spatially constant, but with arbitrary time dependence electric field, parallel to the direction of motion of the particles. The state of these particles remains also unaltered in the presence of a plane electromagnetic wave, for example, a laser beam propagating in a direction opposite to the direction of motion of the particles. The main concept of this work is shown schematically in figure 1.

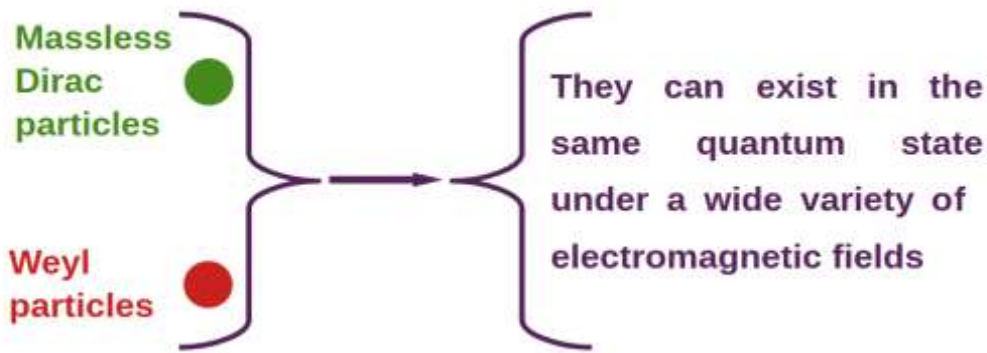

**Figure 1**: The main concept of this work.

Furthermore, in Section 3, we present a general form of solutions to the Weyl equation, which can be utilized to fully control the quantum state of Weyl particles using appropriate electromagnetic fields. These results are expected to play an important role in the practical applications of Weyl materials [7-9].

In more detail, it is well known that power dissipation is the main obstacle in modern electronics. Consequently, the property of Weyl particles to conduct electricity much faster and more efficiently compared to normal electrons [7] is very attractive from a practical point of view and could be utilized to develop novel, more efficient devices for use in nanoelectronics, such as ultrafast switches, spin-transistors, logic devices, electric and magnetic field sensors, etc. In addition, Hills and collaborators [9] have shown that an heterostructure made of two Weyl semimetals, such as NbAs and NbP, exhibits a negative refractive index behavior. Based on this conclusion, it was shown [9] that it is possible to design a three-dimensional Veselago lens made of three layers of different Weyl materials, which could be utilized as a probing tip for scanning tunneling microscopes (STMs). An STM designed in this way could focus electron beams onto sub-angstrom regions, boosting precision to levels at which the technique could routinely see individual atomic orbitals and chemical bonds. Obviously, further design, development and materialization of these, and potentially many more devices and applications, require a deeper understanding of the electromagnetic interactions of Weyl particles, offered by the present and other studies. The present work has also the advantage of providing a means to fully control the state of Weyl particles through appropriate electromagnetic fields, which could be proven particularly useful for the further development of novel devices and applications.

Finally, in Section 4, we study the transition from degenerate to non-degenerate solutions as the particles acquire mass. It is shown that, as far as the rest energy of the particles is much smaller than their total energy, the concept of the degeneracy can also be extended to massive particles, in an approximate sense, provided that the magnitude of the electromagnetic fields corresponding to the exact degenerate solutions is sufficiently small.

## 2. General form of degenerate solutions to the Dirac equation for massless particles and the corresponding electromagnetic fields

In this section we provide a general form of the degenerate solutions to the Dirac equation for massless particles. Furthermore, we will calculate the electromagnetic fields corresponding to these solutions, providing some examples regarding both spatially constant electromagnetic fields and electromagnetic waves.

In a recent article [10], we showed that all solutions to the Dirac equation,

$$i\gamma^{\mu}\partial_{\mu}\Psi + a_{\mu}\gamma^{\mu}\Psi - m\Psi = 0 \tag{1}$$

satisfying the conditions $\Psi^{\dagger}\gamma\,\Psi = 0$ and $\Psi^{T}\gamma^{2}\,\Psi \neq 0$, are degenerate, corresponding to an infinite number of electromagnetic 4-potentials, which are explicitly calculated in Theorem 5.4 in [10]. Here, $\gamma^{\mu}$ are the standard Dirac matrices, $\gamma = \gamma^{0} + i\gamma^{1}\gamma^{2}\gamma^{3}$, $m$ is the mass of the particle and $a_{\mu} = qA_{\mu}$, where $q$ is the electric charge of the particle, and $A_{\mu}$ is the electromagnetic 4-potential. It should also be noted that Eq. (1) is written in natural units, where $\hbar = c = 1$.

In an effort to find general forms of degenerate solutions to the Dirac equation, we found that all spinors of the form

$$\Psi_p = (c_1 u_\uparrow + c_2 u_\downarrow) \exp[ih(x,y,z,t)]$$
$$\Psi_a = (c_1 v_\uparrow + c_2 v_\downarrow) \exp[ih(x,y,z,t)] \quad (2)$$

are degenerate solutions to the Dirac equation corresponding to massless particles or antiparticles propagating along a direction in space defined by the angles $(\theta, \varphi)$ in spherical coordinates. Here, $c_1, c_2$ are arbitrary complex constants, $h$ is an arbitrary real function of the spatial coordinates and time, and $u_\uparrow, u_\downarrow, v_\uparrow, v_\downarrow$ are eigenvectors describing the spin state of the particle $(u_\uparrow, u_\downarrow)$ or antiparticle $(v_\uparrow, v_\downarrow)$. In the case of massless particles, they are given by the formulae [11]:

$$u_\uparrow = \left( \cos\frac{\theta}{2} \quad e^{i\varphi}\sin\frac{\theta}{2} \quad \cos\frac{\theta}{2} \quad e^{i\varphi}\sin\frac{\theta}{2} \right)^T$$

$$u_\downarrow = \left( -\sin\frac{\theta}{2} \quad e^{i\varphi}\cos\frac{\theta}{2} \quad \sin\frac{\theta}{2} \quad -e^{i\varphi}\cos\frac{\theta}{2} \right)^T \quad (3)$$

$$v_\uparrow = \left( \sin\frac{\theta}{2} \quad -e^{i\varphi}\cos\frac{\theta}{2} \quad -\sin\frac{\theta}{2} \quad e^{i\varphi}\cos\frac{\theta}{2} \right)$$

$$v_\downarrow = \left( \cos\frac{\theta}{2} \quad e^{i\varphi}\sin\frac{\theta}{2} \quad \cos\frac{\theta}{2} \quad e^{i\varphi}\sin\frac{\theta}{2} \right)$$

The 4-potentials corresponding to the above solutions are

$$(a_0, a_1, a_2, a_3) = \left( \frac{\partial h}{\partial t}, \frac{\partial h}{\partial x}, \frac{\partial h}{\partial y}, \frac{\partial h}{\partial z} \right) \quad (4)$$

Further, according to Theorem 5.4 in [10], spinors (3) will also be solutions to the Dirac equation for an infinite number of 4-potentials, given by the formula

$$b_\mu = a_\mu + s\kappa_\mu \quad (5)$$

where

$$(\kappa_0, \kappa_1, \kappa_2, \kappa_3) = \left( 1, -\frac{\Psi^T \gamma^0 \gamma^1 \gamma^2 \Psi}{\Psi^T \gamma^2 \Psi}, -\frac{\Psi^T \gamma^0 \Psi}{\Psi^T \gamma^2 \Psi}, \frac{\Psi^T \gamma^0 \gamma^2 \gamma^3 \Psi}{\Psi^T \gamma^2 \Psi} \right)$$
$$= (1, -\sin\theta\cos\varphi, -\sin\theta\sin\varphi, -\cos\theta) \quad (6)$$

and $s$ is an arbitrary real function of the spatial coordinates and time.

The electromagnetic fields (in Gaussian units) corresponding to the above 4-potetials are [12]

$$\mathbf{E} = -\nabla U - \frac{\partial \mathbf{A}}{\partial t} = -\left(\sin\theta\cos\varphi \frac{\partial s_q}{\partial t} + \frac{\partial s_q}{\partial x}\right)\mathbf{i}$$

$$-\left(\sin\theta\sin\varphi \frac{\partial s_q}{\partial t} + \frac{\partial s_q}{\partial y}\right)\mathbf{j} - \left(\cos\theta \frac{\partial s_q}{\partial t} + \frac{\partial s_q}{\partial z}\right)\mathbf{k}$$

$$\mathbf{B} = \nabla \times \mathbf{A} = \left(-\sin\theta\sin\varphi \frac{\partial s_q}{\partial z} + \cos\theta \frac{\partial s_q}{\partial y}\right)\mathbf{i} + \left(\sin\theta\cos\varphi \frac{\partial s_q}{\partial z} - \cos\theta \frac{\partial s_q}{\partial x}\right)\mathbf{j}$$

$$+ \sin\theta\left(-\cos\varphi \frac{\partial s_q}{\partial y} + \sin\varphi \frac{\partial s_q}{\partial x}\right)\mathbf{k}$$

(7)

Here $U = b_0/q$ is the electric potential and $\mathbf{A} = -(1/q)(b_1\mathbf{i} + b_2\mathbf{j} + b_3\mathbf{k})$ is the magnetic vector potential, where $s_q = s/q$. The above equations are also written in the natural system of units, where $\hbar = c = 1$. The minus sign in the definition of the magnetic potential is related to the form of the Dirac equation used in this article.

Since the choice of the coordinate system is arbitrary, the direction of motion of the particles can be set to correspond to the $+z-direction$ without loss of generality. In this case $\theta = \varphi = 0$, and equations (7) obtain the simplified form:

$$\mathbf{E} = -\frac{\partial s_q}{\partial x}\mathbf{i} - \frac{\partial s_q}{\partial y}\mathbf{j} - \left(\frac{\partial s_q}{\partial z} + \frac{\partial s_q}{\partial t}\right)\mathbf{k}$$

$$\mathbf{B} = \frac{\partial s_q}{\partial y}\mathbf{i} - \frac{\partial s_q}{\partial x}\mathbf{j}$$

(8)

Thus, the state of the particle (or antiparticle) remains unchanged under a wide variety of electromagnetic fields. For example, assuming that the arbitrary function $s_q$ depends only on time, Eq. (8) implies that the state of the particle will not change in the presence of a spatially constant, but with an arbitrary time-dependence, electric field applied along the direction of motion of the particle.

Further, assuming that the arbitrary function $s_q$ is given by the formula

$$s_q = -E_{W1}\cos\left[k_W(z-t) + \delta_{W1}\right]x - E_{W2}\cos\left[k_W(z-t) + \delta_{W2}\right]y \tag{9}$$

where $E_{W1}, k_W, \delta_{W1}, E_{W2}, \delta_{W2}$ are real constants, the electromagnetic fields (8) take the form:

$$\mathbf{E} = \mathbf{E}_W, \qquad \mathbf{B} = \mathbf{B}_W \tag{10}$$

where

$$\mathbf{E}_W = E_{W1}\cos\left[k_W(z-t) + \delta_{W1}\right]\mathbf{i} + E_{W2}\cos\left[k_W(z-t) + \delta_{W2}\right]\mathbf{j}$$

$$\mathbf{B}_W = -E_{W2}\cos\left[k_W(z-t) + \delta_{W2}\right]\mathbf{i} + E_{W1}\cos\left[k_W(z-t) + \delta_{W1}\right]\mathbf{j}$$

(11)

The above electromagnetic fields correspond to an electromagnetic wave, for example, a laser beam, of arbitrary polarization propagating along the $+z-direction$ with a Poynting vector

$$\mathbf{S} = \frac{1}{4\pi} \mathbf{E}_W \times \mathbf{B}_W$$
$$= \frac{1}{4\pi} \left[ E_{W1}^2 \cos^2\left[k_W(z-t)+\delta_{W1}\right] + E_{W2}^2 \cos^2\left[k_W(z-t)+\delta_{W2}\right] \right] \mathbf{k} \tag{12}$$

Thus, the state of massless Dirac particles described by spinors of the general form given by Eq. (2) will not be affected by the presence of a plane electromagnetic wave (e.g., a laser beam) of arbitrary polarization propagating in a direction parallel to the direction of motion of the particles.

Another important remark is that, setting $h = E(z-t)$ in Eq. (2a), and $h = -E(z-t)$ in Eq. (2b), we obtain the familiar form:

$$\Psi_p = (c_1 u_\uparrow + c_2 u_\downarrow) \exp\left[iE(z-t)\right]$$
$$\Psi_a = (c_1 v_\uparrow + c_2 v_\downarrow) \exp\left[-iE(z-t)\right] \tag{13}$$

describing free massless particles or antiparticles with energy $E$ moving along the $+z-direction$.

3. **General forms of solutions to the Weyl equation and a proposed method to fully control the quantum state of Weyl particles using appropriate electromagnetic fields**

In this section, some general solutions to the Weyl equation are provided and the corresponding electromagnetic fields are calculated. Based on these results, we propose a method for fully controlling the quantum states of Weyl particles, using appropriate electromagnetic fields.

It can be easily observed that if the coefficients $c_1$ or $c_2$ in Eq. (2) are equal to zero, the quantity $\Psi^T \gamma_2 \Psi$ also becomes zero. In this case, considering Theorem 5.4 in [10], the solution given by Eq. (2) can be written in the form $\Psi = (\psi, \psi)^T$ or $\Psi = (\psi, -\psi)^T$, where $\psi$ is a solution to the Weyl equation. In the case of particles with spin parallel to their direction of motion (positive helicity) the Weyl equation can be written in the form

$$i\sigma^\mu \partial_\mu \Psi + a_\mu \sigma^\mu \Psi = 0 \tag{14}$$

while in the case of particles with spin antiparallel to their direction of motion (negative helicity) it can be written in the form

$$i\sigma^\mu \partial_\mu \Psi - 2i\sigma^0 \partial_0 \Psi + a_\mu \sigma^\mu \Psi - 2a_0 \sigma^0 \Psi = 0 \tag{15}$$

where $\sigma^\mu$ are the standard Pauli matrices, $a_\mu = qA_\mu$, $q$ is the electric charge of the particles and $A_\mu$ is the electromagnetic 4-potential, as in the case of the Dirac equation. Equations (14) and (15) can also describe antiparticles with negative and positive helicities, respectively.

Specifically, all spinors of the form

$$\psi_1 = \begin{pmatrix} \cos\left(\dfrac{\theta}{2}\right) \\ e^{i\varphi} \sin\left(\dfrac{\theta}{2}\right) \end{pmatrix} \exp\left[ih(x,y,z,t)\right] \qquad (16)$$

are solutions to the Weyl equation (14) for the real 4-potential given by Eq. (4). In addition, all spinors of the form

$$\psi_2 = \begin{pmatrix} -\sin\left(\dfrac{\theta}{2}\right) \\ e^{i\varphi} \cos\left(\dfrac{\theta}{2}\right) \end{pmatrix} \exp\left[ih(x,y,z,t)\right] \qquad (17)$$

are solutions to the Weyl equation (15) for the same 4-potential. Further, according to Theorem 3.1 in [10], the spinors given by Eq. (16) will also be solutions to the Weyl equation (14) for the 4-potentials

$$b_{\mu 1} = a_{\mu 1} + s\kappa_{\mu 1} \qquad (18)$$

where

$$\begin{aligned}(\kappa_{01}, \kappa_{11}, \kappa_{21}, \kappa_{31}) &= \left(1, -\dfrac{\psi_1^\dagger \sigma^1 \psi_1}{\psi_1^\dagger \psi_1}, -\dfrac{\psi_1^\dagger \sigma^2 \psi_1}{\psi_1^\dagger \psi_1}, -\dfrac{\psi_1^\dagger \sigma^3 \psi_1}{\psi_1^\dagger \psi_1}\right) \\ &= (1, -\sin\theta\cos\varphi, -\sin\theta\sin\varphi, -\cos\theta)\end{aligned} \qquad (19)$$

and $s$ is an arbitrary real function of spatial coordinates and time. Similarly, the spinors given by Eq. (17) will also be solutions to the Weyl equation (15) for the 4-potentials

$$b_{\mu 2} = a_{\mu 2} + s\kappa_{\mu 2} \qquad (20)$$

where

$$\begin{aligned}(\kappa_{02}, \kappa_{12}, \kappa_{22}, \kappa_{32}) &= \left(1, \dfrac{\psi_2^\dagger \sigma^1 \psi_2}{\psi_2^\dagger \psi_2}, \dfrac{\psi_2^\dagger \sigma^2 \psi_2}{\psi_2^\dagger \psi_2}, \dfrac{\psi_2^\dagger \sigma^3 \psi_2}{\psi_2^\dagger \psi_2}\right) \\ &= (1, -\sin\theta\cos\varphi, -\sin\theta\sin\varphi, -\cos\theta)\end{aligned} \qquad (21)$$

From the above analysis it is clear that $a_{\mu 1} = a_{\mu 2}$ and $\kappa_{\mu 1} = \kappa_{\mu 2}$. Consequently, $b_{\mu 1} = b_{\mu 2}$. In addition, the 4-potentials $b_{\mu 1}$, $b_{\mu 2}$ are identical to those given by Eq. (5), corresponding to massless Dirac particles. Thus, the state of a Weyl particle described by the spinors (16), (17) will not be affected by the wide variety of electromagnetic fields given by Eq. (7). Furthermore, all the physical properties regarding the electromagnetic interactions of Weyl particles described by the spinors (16), (17), will be the same as those of massless Dirac particles, as discussed in section 2. This is expected because Weyl particles can be considered as massless Dirac particles with either positive or negative helicity.

An interesting remark regarding both Dirac and Weyl equations is that the spinors resulting from multiplying the solutions (2), (16) and (17) with an arbitrary real function $d(w)$ where

$$w = x\sin\theta\cos\varphi + y\sin\theta\sin\varphi x + z\cos\theta - t \qquad (22)$$

are also solutions to these equations for the same 4-potentials. However, in this case, the 4-vector probability current $j^\mu$, defined as:

$$j^\mu = \begin{pmatrix} \Psi^\dagger\Psi \\ \Psi^\dagger\gamma^0\gamma^1\Psi \\ \Psi^\dagger\gamma^0\gamma^2\Psi \\ \Psi^\dagger\gamma^0\gamma^3\Psi \end{pmatrix} \qquad (23)$$

in the case of the Dirac equation, and similarly in the case of the Weyl equation, becomes proportional to $d^2$. Consequently, the continuity equation [11]

$$\partial_\mu j^\mu = 0 \qquad (24)$$

is not generally satisfied, except for the special case of $d$ being a constant. For example, setting

$$d = \exp(-\lambda w^2) \qquad (25)$$

where $\lambda$ is a positive real constant, we obtain that $j^\mu \to 0$ in the case that

$$|w| \to \infty \qquad (26)$$

This practically means that these solutions behave in an unusual way, existing only at a specific "region" of space and time, and disappearing as $|w|$ tends to infinity. The nature of these solutions and their connection to the physical reality is an interesting open question that could be investigated in a future work.

Another particularly important remark regarding Weyl particles is that the solutions given by equations (16) and (17) can be generalized, allowing the angles $(\theta,\varphi)$ to be functions of time. Thus, the spinors

$$\psi_1 = \begin{pmatrix} \cos\left(\dfrac{\theta(t)}{2}\right) \\ e^{i\varphi(t)} \sin\left(\dfrac{\theta(t)}{2}\right) \end{pmatrix} \exp\left[ih(x,y,z,t)\right] \qquad (27)$$

$$\psi_2 = \begin{pmatrix} -\sin\left(\dfrac{\theta(t)}{2}\right) \\ e^{i\varphi(t)} \cos\left(\dfrac{\theta(t)}{2}\right) \end{pmatrix} \exp\left[ih(x,y,z,t)\right] \qquad (28)$$

are solutions to the Weyl equations (14) and (15) for the 4-potentials

$$\left(a_0, a_1, a_2, a_3\right) = \left(\frac{\partial h}{\partial t} + \frac{1}{2}\frac{d\varphi}{dt},\ \frac{\partial h}{\partial x} + \frac{1}{2}\sin\varphi\frac{d\theta}{dt},\ \frac{\partial h}{\partial y} - \frac{1}{2}\cos\varphi\frac{d\theta}{dt},\ \frac{\partial h}{\partial z} - \frac{1}{2}\frac{d\varphi}{dt}\right) \qquad (29)$$

and

$$\left(a'_0, a'_1, a'_2, a'_3\right) = \left(\frac{\partial h}{\partial t} + \frac{1}{2}\frac{d\varphi}{dt},\ \frac{\partial h}{\partial x} - \frac{1}{2}\sin\varphi\frac{d\theta}{dt},\ \frac{\partial h}{\partial y} + \frac{1}{2}\cos\varphi\frac{d\theta}{dt},\ \frac{\partial h}{\partial z} + \frac{1}{2}\frac{d\varphi}{dt}\right) \qquad (30)$$

respectively. In addition, according to Theorem 3.1 in [10], the spinors (16) and (17) will also be solutions to the Weyl equations (14) and (15) for the 4-potentials

$$b_\mu = a_\mu + \kappa_\mu s, \qquad b'_\mu = a'_\mu + \kappa_\mu s \qquad (31)$$

respectively, where

$$\left(\kappa_0, \kappa_1, \kappa_2, \kappa_3\right) = \left(1, -\sin\theta(t)\cos\varphi(t), -\sin\theta(t)\sin\varphi(t), -\cos\theta(t)\right) \qquad (32)$$

In both cases, $s$ is an arbitrary real function of spatial coordinates and time.

This result is particularly important because it provides a method to control the quantum state of a Weyl particle by applying appropriate electromagnetic fields. Specifically, setting the function $h$ in Eqs. (16), (17) as

$$h(x,y,z,t) = E_0\left[\sin\theta(t)\cos\varphi(t)\,x + \sin\theta(t)\sin\varphi(t)\,y + \cos\theta(t)\,z - t\right] \qquad (33)$$

the resulting spinors describe a particle with energy $E_0$ and propagation direction, defined by the angles $\theta(t), \varphi(t)$ in spherical coordinates, with $\theta(t)$ being the polar and $\varphi(t)$ the azimuthal angle. The electromagnetic fields corresponding to these solutions are:

$$\mathbf{E} = \frac{1}{2q}\left(\cos\varphi \frac{d\theta}{dt}\frac{d\varphi}{dt} + \sin\varphi \frac{d^2\theta}{dt^2}\right)\mathbf{i}$$
$$+ \frac{1}{2q}\left(\sin\varphi \frac{d\theta}{dt}\frac{d\varphi}{dt} - \cos\varphi \frac{d^2\theta}{dt^2}\right)\mathbf{j} \quad (34)$$
$$- \frac{1}{2q}\frac{d^2\varphi}{dt^2}\mathbf{k}$$
$$\mathbf{B} = \mathbf{0}$$

in the case of particles with positive helicity and $\mathbf{E'} = -\mathbf{E}$, $\mathbf{B'} = \mathbf{0}$ in the case of particles with negative helicity. The application of the above electromagnetic fields allows the determination of the propagation direction of the Weyl particles as function of time. The remarkable properties of the solutions (16), (17) are extensively studied in [13], where it is shown that they can describe Weyl particles existing at different states in zero electromagnetic field. They can also describe localized states of Weyl particles, leading to the generation of mass through the localization of energy in Einstein's field equations of general relativity [13].

Furthermore, the state of the particles remains unaffected when adding the following electromagnetic fields to the ones given by Eq. (34):

$$\mathbf{E}_s(\mathbf{r},t) = -\frac{1}{q}\left[\sin\theta\cos\varphi \frac{\partial s}{\partial t} + \frac{\partial s}{\partial x} + s\left(\cos\theta\cos\varphi \frac{d\theta}{dt} - \sin\theta\sin\varphi \frac{d\varphi}{dt}\right)\right]\mathbf{i}$$
$$-\frac{1}{q}\left[\sin\theta\sin\varphi \frac{\partial s}{\partial t} + \frac{\partial s}{\partial y} + s\left(\cos\theta\sin\varphi \frac{d\theta}{dt} + \sin\theta\cos\varphi \frac{d\varphi}{dt}\right)\right]\mathbf{j}$$
$$-\frac{1}{q}\left(\cos\theta \frac{\partial s}{\partial t} + \frac{\partial s}{\partial z} + \sin\theta \frac{d\theta}{dt} s\right)\mathbf{k} \quad (35)$$
$$\mathbf{B}_s(\mathbf{r},t) = \frac{1}{q}\left(-\sin\theta\sin\varphi \frac{\partial s}{\partial z} + \cos\theta \frac{\partial s}{\partial y}\right)\mathbf{i} + \frac{1}{q}\left(\sin\theta\cos\varphi \frac{\partial s}{\partial z} - \cos\theta \frac{\partial s}{\partial x}\right)\mathbf{j}$$
$$+\frac{1}{q}\sin\theta\left(-\cos\varphi \frac{\partial s}{\partial y} + \sin\varphi \frac{\partial s}{\partial x}\right)\mathbf{k}$$

Consequently, equations (34) and (35) describe the family of electromagnetic fields that should be applied to control the state of Weyl particles regarding their propagation direction $(\theta(t),\varphi(t))$ as function of time. The energy of the particles can be controlled through the arbitrary function $s(\mathbf{r},t)$, as shown in [13]. In that work, we also provide a thorough analysis regarding the control of the localization of Weyl particles through appropriate electromagnetic fields, showing that the localization of the energy of the particles can lead to the appearance of gravitational mass through Einstein's field equations of general relativity.

As it is also mentioned in the introduction, this result is expected to play a particularly important role in the development of novel devices and applications using materials supporting Weyl fermions [2, 5-9].

## 4. A discussion on the transition from degenerate to non-degenerate solutions as the particles acquire mass

In this section, we discuss the transition from degenerate to non-degenerate solutions as the particles acquire mass and conclude that in an approximate sense, our results can also be extended to the case of massive particles, provided that the rest energy of the particles is much smaller than their total energy and the magnitude of the electromagnetic fields corresponding to the exact degenerate solutions is sufficiently small.

In the case of massive particles, the eigenvectors given by equations (3), describing the spin state of the particle, become [11]:

$$u_\uparrow = \begin{pmatrix} \cos\left(\frac{\theta}{2}\right) \\ e^{i\varphi}\sin\left(\frac{\theta}{2}\right) \\ \frac{|\mathbf{p}|}{E+m}\cos\left(\frac{\theta}{2}\right) \\ \frac{|\mathbf{p}|}{E+m}e^{i\varphi}\sin\left(\frac{\theta}{2}\right) \end{pmatrix} \quad u_\downarrow = \begin{pmatrix} -\sin\left(\frac{\theta}{2}\right) \\ e^{i\varphi}\cos\left(\frac{\theta}{2}\right) \\ \frac{|\mathbf{p}|}{E+m}\sin\left(\frac{\theta}{2}\right) \\ -\frac{|\mathbf{p}|}{E+m}e^{i\varphi}\cos\left(\frac{\theta}{2}\right) \end{pmatrix} \quad (36)$$

$$v_\uparrow = \begin{pmatrix} \frac{|\mathbf{p}|}{E+m}\sin\left(\frac{\theta}{2}\right) \\ -\frac{|\mathbf{p}|}{E+m}e^{i\varphi}\cos\left(\frac{\theta}{2}\right) \\ -\sin\left(\frac{\theta}{2}\right) \\ e^{i\varphi}\cos\left(\frac{\theta}{2}\right) \end{pmatrix} \quad v_\downarrow = \begin{pmatrix} \frac{|\mathbf{p}|}{E+m}\cos\left(\frac{\theta}{2}\right) \\ \frac{|\mathbf{p}|}{E+m}e^{i\varphi}\sin\left(\frac{\theta}{2}\right) \\ \cos\left(\frac{\theta}{2}\right) \\ e^{i\varphi}\sin\left(\frac{\theta}{2}\right) \end{pmatrix} \quad (37)$$

where $|\mathbf{p}| = \sqrt{E^2 - m^2}$ is the modulus of the momentum of the particle in the natural system of units $(\hbar = c = 1)$. Substituting the above eigenvectors into the spinors given by Eq. (2), it can be observed that they continue to satisfy the massless Dirac equation for the 4-potentials described by Eq. (4). However, these spinors are no longer degenerate, because $\Psi^\dagger \gamma \Psi \neq 0$. Specifically, it was found that in the case of particles the following holds: $\Psi_p^\dagger \gamma \Psi_p = 2e(|c_1|^2 + |c_2|^2)/(1+e)$, where $e = m/E$. Similarly, in the case of antiparticles: $\Psi_a^\dagger \gamma \Psi_a = -2e(|c_1|^2 + |c_2|^2)/(1+e)$. However, if the rest energy of the particle is much smaller than its total energy, the parameter $e$ tends to zero. In this case the spinors given by Eq. (2) with the "massive" eigenvectors defined by equations (36), (37) can be considered as nearly degenerate, because $\Psi_p^\dagger \gamma \Psi_p$ and $\Psi_a^\dagger \gamma \Psi_a$ tend to zero.

However, the aforementioned spinors are no longer solutions to the massless Dirac equation for the 4-potentials given by Eq. (5). Indeed, substituting these spinors into the massless Dirac equation using the 4-potentials (5), we obtain that

$$i\gamma^\mu \partial_\mu \Psi_p + b_\mu \gamma^\mu \Psi_p = \left(1 - \sqrt{\frac{1-e}{1+e}}\right) s \Psi_p^* \tag{38}$$

$$i\gamma^\mu \partial_\mu \Psi_a + b_\mu \gamma^\mu \Psi_a = -\left(1 - \sqrt{\frac{1-e}{1+e}}\right) s \Psi_a^*, \tag{39}$$

Where $\Psi_p^*$, $\Psi_a^*$ are the unperturbed spinors for particles and antiparticles respectively. However, in the case that the rest energy of the particles is much smaller than their total energy $(e \ll 1)$, the above equations take the simpler form

$$i\gamma^\mu \partial_\mu \Psi_p + b_\mu \gamma^\mu \Psi_p = e\, s \Psi_p^* \tag{40}$$

$$i\gamma^\mu \partial_\mu \Psi_a + b_\mu \gamma^\mu \Psi_a = -e\, s \Psi_a^* \tag{41}$$

This result implies that as the ratio $e$ of the rest energy to the total energy of the particles increases, the function $s$ should be restricted to smaller values, suppressing the effects of degeneracy. On the other hand, as the ratio $e$ decreases, the function $s$ is allowed to take larger values, and the degeneracy becomes more evident. Finally, as the ratio $e$ tends to zero, there is no restriction on the values of the function $s$ and the theory of degeneracy becomes fully applicable.

## 5. Conclusions

In conclusion, we have found general forms of degenerate solutions to the massless Dirac and Weyl equations and calculated the electromagnetic fields corresponding to these solutions. Specifically, we have shown that the state of massless Dirac particles, described by degenerate spinors, as well as Weyl particles, is not affected by the presence of a spatially constant, but with arbitrary time dependence electric field applied along the direction of motion of the particles, or a plane electromagnetic wave, such as a laser beam propagating in a direction opposite to the direction of motion of the particles. Furthermore, based on a special class of solutions to the Weyl equation, we proposed a method for fully controlling the quantum state of Weyl particles using appropriate electromagnetic fields. These results are expected to play a particularly important role in the development of novel devices and applications using materials supporting massless Dirac and/or Weyl particles.

Finally, we studied the transition from degenerate to non-degenerate solutions as the particles acquire mass, showing that the concept of degeneracy is still valid in an approximate sense, provided that the rest energy of the particles is much smaller than their total energy and the magnitude of the electromagnetic fields corresponding to the exact degenerate solutions is sufficiently small.